\documentclass{PoS}
\newcommand{\tm}[1]{\mathrm{ #1}}

\title{Holography for Heavy Quarks and Mesons at Finite Chemical Potential}

\ShortTitle{Holography for Heavy Quarks and Mesons at Finite Chemical Potential}

\author{\speaker{Carlo Ewerz}, Ling Lin, Andreas Samberg, Konrad Schade\\
        Institut f\"ur Theoretische Physik, Universit\"at Heidelberg,\\ 
        Philosophenweg 16, 69120 Heidelberg, Germany\\
        and\\
        ExtreMe Matter Institute EMMI, GSI Helmholtzzentrum f\"ur Schwerionenforschung\\
        Planckstra{\ss}e 1, 64291 Darmstadt, Germany\\
        E-mail: \email{C.Ewerz@thphys.uni-heidelberg.de}, \email{L.Lin@thphys.uni-heidelberg.de}, \email{A.Samberg@thphys.uni-heidelberg.de}, \email{K.Schade@thphys.uni-heidelberg.de}}

\abstract{
We study the properties of heavy quarks as probes of strongly coupled plasmas 
with and without chemical potential by means of the gauge/gravity (AdS/CFT) duality. 
We compute the screening distance of a heavy quark-antiquark pair, its free energy, 
and the running coupling in large classes of non-conformal models arising as deformations of 
pure AdS space. 
We further investigate the energy loss of a heavy quark moving on a circular orbit 
as an example of an accelerated motion. 
These observables exhibit universal features 
independent of the deformation, pointing to strong-coupling universality. 
Our results should be relevant for processes involving heavy quarks and their 
bound states in the quark-gluon plasma, including the case of finite net baryon density.
}

\FullConference{9th International Workshop on Critical Point and Onset of Deconfinement - CPOD2014,\\
		17-21 November 2014\\
		ZiF (Center of Interdisciplinary Research), University of Bielefeld, Germany}

\begin{document}

\section{Introduction}
\label{sec:intro}

The AdS/CFT (or gauge/gravity) duality 
\cite{Maldacena:1997re,Gubser:1998bc,Witten:1998qj} 
offers the possibility to address a variety of questions 
concerning strongly coupled systems in a holographic setup. 
One of the most 
interesting applications of the duality is the study of the 
strongly coupled quark-gluon plasma produced in heavy-ion 
collisions at RHIC and LHC, for recent reviews see for example 
\cite{CasalderreySolana:2011us,DeWolfe:2013cua}. 
While static properties of the quark-gluon plasma can 
be well described in lattice gauge theory, dynamical phenomena 
at strong coupling pose a more difficult challenge for which 
the AdS/CFT duality is particularly well suited. 

Originally, the AdS/CFT duality was found to relate the dynamics 
of maximally supersymmetric $\mbox{SU}(N_c)$ Yang-Mills 
theory (SYM) in the large-$N_c$ limit and at strong coupling to 
weakly coupled gravity on 5-dimensional Anti-de Sitter (AdS${}_5$) space. 
Loosely speaking, one can think of the gauge theory as living on 
the 4-dimensional boundary of the 5-dimensional AdS space, 
that is at $z=0$ where $z$ denotes the additional holographic (bulk) 
coordinate. A finite temperature of the gauge theory is on the gravity side 
of the duality represented by a black-hole horizon in the bulk located at 
a point $z_h$ in the fifth coordinate and extended in the other 
4 space-time directions. In addition, a chemical potential of the 
gauge theory can be described in this setup by giving the black hole 
an electric charge. 

Maximally supersymmetric Yang-Mills theory, a scale invariant or conformal 
field theory (hence the name CFT) differs strongly from QCD. 
The applicability of the AdS/CFT duality to real-world processes is therefore 
far from obvious. However, at finite temperature the two theories appear less different: 
On the one hand, QCD enters a deconfined phase without chiral 
symmetry breaking above the critical temperature and many observables 
become almost scale invariant. In SYM theory, on the other hand, supersymmetry 
and conformal symmetry are broken at finite 
temperature. But this is not really sufficient for making reliable predictions 
based on the AdS/CFT duality. More promising is the construction of 
dual gravity theories in which conformal symmetry is broken from the outset. 
This requires to introduce some sort of deformation of the AdS space 
of the original duality as we will describe in more detail below. 
In one important line of research one tries to construct theories of this 
type which model many properties of QCD as precisely as possible, see 
for example \cite{Gursoy:2010fj}, although an exact gravity dual of QCD 
seems prohibitively difficult to find in this way. We follow a different 
approach here by looking at large classes of deformations of AdS space. 
It turns out that some observables exhibit a remarkable universality in 
large classes of theories, either by having a fixed value independent of 
the deformation, or by changing only very little or consistently in one 
direction as the theory is deformed. 
The most prominent example of such an observable is the ratio $\eta/s$ of 
shear viscosity to entropy density which assumes the value $1/(4 \pi)$ in 
many theories obtained as duals of deformed AdS spaces 
\cite{Policastro:2001yc,Kovtun:2004de}. Universal properties of this 
kind might be a sign of universal strong-coupling dynamics that is 
more or less independent of the underlying microscopic theory. 
If indeed universal in this sense, such behavior would have a good chance 
to be very relevant for the study of the quark-gluon plasma explored in 
experiments. 

\section{Holographic Models}
\label{sec:models}

Let us now discuss in more detail the relevant AdS spaces and 
their non-conformal deformations which describe strongly coupled 
theories at finite temperature and at finite chemical potential. 

\subsection{Finite Temperature}
\label{subsec:fintemp}

According to the AdS/CFT correspondence, $\mathcal{N}\!=\!4$ 
supersymmetric Yang-Mills 
theory at finite temperature is dual to gravity on AdS${}_5$ with a Schwarzschild 
black hole at a position $z_h$ in the bulk, 
\begin{equation}
\label{n4metric}
\tm{d}s^2 = \frac{L_{\rm AdS}^2}{z^2} \, \left[ - f(z) \, \tm{d}t^2 
+ \tm{d} \vec{x}\,{}^2 + \frac{\tm{d}z^2}{f(z)} \right] \,,
\end{equation}
where 
\begin{equation}
\label{n4h}
f(z)= 1 - \frac{z^4}{z_h^4} \,.
\end{equation}
$L_{\rm AdS}$, the AdS curvature radius, is assumed to be large which 
implies a large 't Hooft coupling of the gauge theory. The temperature of the 
gauge theory coincides with the Hawking temperature of the black hole 
and is given by $T=(\pi z_h)^{-1}$. 

Theories closer to actual QCD can be obtained by deforming the AdS metric. 
One possibility is to modify the above metric in an ad-hoc way by introducing 
a simple factor which breaks conformal invariance. 
In \cite{Andreev:2006ct} and \cite{Kajantie:2006hv} such a deformation 
was proposed in which the metric (\ref{n4metric}) is multiplied by an 
overall $z$-dependent factor $\exp(c z^2)$ with a dimensionful 
parameter $c$. The relation between temperature and the position 
of the horizon, $T=(\pi z_h)^{-1}$, is not changed. 
The resulting model is a finite-temperature version of 
the popular soft-wall model at $T=0$ \cite{Karch:2006pv}, 
and we hence denote it by SW${}_T$. 
This model allows to obtain many interesting quantities 
in simple calculations, often even analytically. But the model 
does not solve the equations of motion of any 5-dimensional 
gravitational Einstein-Hilbert action. As a consequence, its 
consistency is questionable. For example, one does not expect 
it to lead to consistent thermodynamics in the dual 4-dimensional 
theory. The model also gives rise to unphysical solutions for 
certain string configurations corresponding to the motion of 
a heavy quark through the plasma in the dual theory \cite{ELS}. 

A more promising way to make closer contact 
with the quark-gluon plasma is to add further fields to the 5-dimensional 
gravity, and then to obtain a non-conformal metric as a solution 
of a gravitational action. In the simplest case one adds only 
one scalar field $\Phi$ as proposed in \cite{Gubser:2008ny,DeWolfe:2009vs}, 
considering the Einstein-Hilbert action 
\begin{equation}
\label{eq.NonconfMeMo:ConDefMeMo:5DEHdaction}
S_{\rm EHs} = \frac{1}{16 \pi G^{(5)}_\tm{N}} \, \int \tm{d}^5 x \, 
\sqrt{- G} \, \left( \mathcal{R} - \frac{1}{2} ( \partial \Phi )^2 
- V(\Phi) \right) \,,
\end{equation}
where $G$ is the determinant of the deformed metric $G^{\mu\nu}$, 
$G^{(5)}_\tm{N}$ is the 5-dimensional Newton constant, 
$\mathcal{R}$ is the Ricci scalar, and $V(\Phi)$ is a potential 
for the scalar. We are interested in classes of theories in which 
the metric has the general form 
\begin{equation}
\label{eq.ObsPhysQuant:RunCoup:Screen:Metric}
\tm{d} s^2 = \tm{e}^{2 A(z)}  \left( - h(z) \tm{d} t^2 + \tm{d} \vec{x}^{\, 2} \right) 
+ \tm{e}^{2 B(z)} \, \frac{\tm{d} z^2}{h(z)} \,. 
\end{equation}
The resulting temperature of the dual theory is in this case obtained as 
\begin{equation}
\label{eq.NonconfMeMo:SWT:Temp:Tgen}
T = \tm{e}^{A(z_h) - B(z_h)} \, \frac{| h'(z_h)|}{4 \pi} \,,
\end{equation}
with $z_h$ the position of the black hole horizon, that is $h(z_h)=0$. 
The case of the finite-temperature AdS metric (\ref{n4metric}) is recovered 
for $A(z)=B(z)=0$ and $h(z)=f(z)$. 
Metrics obtained as solutions of a 5-dimensional Einstein-Hilbert 
action are consistent in the sense that they will lead to consistent 
thermodynamic relations in the dual theory. At the same time, the holographic 
calculation of observables becomes more complicated if the metric to 
be used first needs to be obtained -- usually numerically -- from 
the equations of motion of (\ref{eq.ObsPhysQuant:RunCoup:Screen:Metric}). 

The construction chosen here gives us a class of bottom-up models for 
plasmas hopefully close to that of QCD. In such bottom-up models, 
we have no a priori constraint on the nature of the scalar field $\Phi$. 
In particular, $\Phi(z)$ can be the dilaton but can also be some other 
type of scalar field. The dilaton typically gives a contribution 
in the action for a string moving on the metric $G^{\mu\nu}$, while 
other scalars will not affect the string in that way.  
We decide to consider both possibilities, taking them as independent models. 
We call the corresponding models 
`string frame' and `Einstein frame' model, respectively. 

The action (\ref{eq.NonconfMeMo:ConDefMeMo:5DEHdaction}) allows one to 
demand solutions resembling models of soft-wall type. Specifically, we can make 
the ansatz 
\begin{equation}
\label{eq.NonconfMeMo:ConDefMeMo:1Para:2ParaMetricAnsatz}
\tm{d} s^2 = \tm{e}^{2 A(\Phi)} \, \big(- h(\Phi) \tm{d} t^2 
+ \tm{d} \vec{x}^{\,2} \big) + \tm{e}^{2 B(\Phi)} \, \frac{\tm{d} \Phi^2}{h(\Phi)} \,. 
\end{equation}
We have used a gauge that identifies the squared holographic coordinate 
with the scalar $\Phi$ where we choose 
\begin{equation}
\label{2pA}
 A(\Phi) = \frac{1}{2} \, \log \left( \frac{L_{\rm AdS}^2}{z^2} \right) - \frac{1}{2} c z^2 \,,
\quad\quad\quad\quad\quad
 \Phi = \sqrt{\frac{3}{2}} \phi z^2 \,,
\end{equation}
and we can then solve the equations of motion 
of (\ref{eq.NonconfMeMo:ConDefMeMo:5DEHdaction}) 
for $B(z)$, the horizon function $h(z)$, 
and the potential $V(\Phi)$. 
To calculate a suitable potential $V(\Phi)$ required 
for a soft-wall type metric might seem unusual from a top-down perspective. 
In our bottom-up approach the main requirement is to have a physically 
interesting class of consistent models, while a particular form of 
the potential $V(\Phi)$ is not necessary. 
As a result of our procedure we have a class of consistent deformed models with 
two independent parameters, $c$ and $\alpha \equiv c/\phi$, cf.\ 
\cite{DeWolfe:2009vs}. Their temperature is obtained via 
(\ref{eq.NonconfMeMo:SWT:Temp:Tgen}) from the 
metric functions at the horizon, that is at the zero of $h$. 

\subsection{Finite Temperature and Finite Chemical Potential}
\label{subsec:finpot}

In the framework of the AdS/CFT duality, $\mathcal{N}\!=\!4$ SYM theory 
with a chemical potential is obtained by making 
the black hole in the holographic dimension charged. The corresponding 
metric is an $AdS_5$ space with a Reissner-Nordstr\"om black hole (AdS-RN). 
It is given by the metric 
\begin{equation}
\label{n4RNmetric}
\tm{d}s^2 = \frac{L_{\rm AdS}^2}{z^2} \, \left[ - h(z) \, \tm{d}t^2 
+ \tm{d} \vec{x}\,{}^2 + \frac{\tm{d}z^2}{h(z)} \right] \,,
\end{equation}
where now 
\begin{equation}
\label{n4RNh}
h(z) = 1 - \left( 1 + Q^2\right) \left(\frac{z}{z_h}\right)^4 
+ Q^2 \left(\frac{z}{z_h}\right)^6 \,.
\end{equation}
The temperature of the boundary theory is given in terms of the charge $Q$ 
of the black hole by $T=(1-\frac{1}{2}Q^2)/(\pi z_h)$. The chemical potential 
is $\mu = \sqrt{3}Q/z_h$, and $Q$ is in the range $0\le Q\le \sqrt{2}$. 
We should point out that the chemical potential implemented in this way is not 
the quark (or baryon) chemical potential of QCD but a chemical potential 
corresponding to the $R$-charge of $\mathcal{N}\!=\!4$ SYM theory. 
However, in this context it can serve as a simple way of introducing 
finite density effects into the system. 

Also in the case of finite chemical potential one can introduce an 
explicit breaking of conformal invariance in order to arrive at a dual 
theory closer to QCD. The simplest way of doing so is again to 
introduce by hand an overall warp factor motivated by the soft-wall model 
\cite{Karch:2006pv}. The corresponding model, obtained by multiplying 
the metric (\ref{n4RNmetric}) by $\exp(\tilde{c}^2 z^2)$, has been proposed and 
studied in \cite{Colangelo:2010pe}. As discussed above, such ad hoc 
models allow one to get a rough picture of the effects of non-conformality, 
but because they do not minimize any gravitational action they suffer from 
inconsistencies. 

Consistent non-conformal metrics dual to theories with chemical potential 
can be found by adding a U(1) field $A_\mu$ to the 5-dimensional 
gravitational action (\ref{eq.NonconfMeMo:ConDefMeMo:5DEHdaction}), 
resulting in the Einstein-Hilbert-Maxwell-scalar action 
\begin{equation}
\label{defAdSRNaction}
S_{\rm EHMs} = \frac{1}{16 \pi G^{(5)}_\tm{N}} \, \int \tm{d}^5 x \, 
\sqrt{- G} \, \left( \mathcal{R} - \frac{1}{2} ( \partial \phi )^2 
- V(\phi) - \frac{f(\phi)}{4} \, F_{\mu\nu}F^{\mu\nu}
\right) \,,
\end{equation}
where $F_{\mu\nu}$ is the field strength of the Maxwell field, and 
$f(\phi)$ determines the coupling of the scalar $\phi$ to $A_\mu$. 
We follow \cite{DeWolfe:2010he} in choosing 
$f(\phi)= \cosh(12/5)/\cosh(6(\phi-2)/5)$ here, but we have checked 
that other choices lead to similar results for our observables. 
Again, we start with the general form (\ref{eq.ObsPhysQuant:RunCoup:Screen:Metric}) 
of the metric. 
For the case of finite chemical potential we make a similar 
but slightly simpler ansatz for $A(z)$ than in the case without chemical 
potential above by setting 
\begin{equation}
\label{ansatz1param}
  A(z) = \log\left(\frac{R}{z}\right),\qquad \phi(z) = \sqrt{\frac{3}{2}} \kappa z^2,\qquad A_\mu\tm{d} x^\mu = \psi(z)\tm{d} t\,.
\end{equation}
Here, $\kappa$ is our dimensionful deformation parameter controlling the 
breaking of conformality. 
Two different classes of deformed models result again from considering 
the scalar $\phi$ as a dilaton or not, again called `string frame' 
and `Einstein frame' model, respectively. The equations of motion 
corresponding to (\ref{defAdSRNaction}) are solved for $B$, $h$, $V$ and $\psi$, and 
the solutions can in fact be found in closed form \cite{SambergDipl}. 
For given $T$ and $\mu$ there is a maximal deformation $\kappa_{\rm max}$ 
for which a black hole solution with these values of $T$ and $\mu$ still exists. 
It turns out that as a consequence a certain region of small $T$ and $\mu$ 
is not accessible with metrics of the chosen form. As discussed above, we 
are for the comparison with the quark-gluon plasma only interested in the 
region of high $T$ and $\mu$ anyway. Therefore this restriction does not 
pose a problem for us. 

\section{Screening Distance and Running Coupling}
\label{sec:screenrun}

Let us now turn to the screening of a heavy quark-antiquark pair 
immersed in a plasma. We allow the pair to move with constant 
velocity $v$ through the plasma. We assume the quarks to be 
infinitely heavy and work in the probe limit. In the AdS/CFT context this 
problem has first been discussed in \cite{Liu:2006nn} for 
$\mathcal{N}\!=\!4$ SYM theory, and in \cite{Liu:2008tz} for the 
case of the ad hoc deformation that we call SW${}_T$. In the AdS 
description the heavy quark-antiquark is represented by an open 
string connecting the two quarks and hanging down into the bulk. 
The ends of the string (i.\,e.\ the quarks) are located at the boundary $z=0$. 
The thermal expectation value of a Wegner-Wilson loop of infinite 
time-extension gives the free energy $F(L)$ of the pair as a function 
of their spatial distance $L$ through 
$\langle W(\mathcal{C}) \rangle = \exp \left[ - i \mathcal{T} \, F(L) \right]$ 
with the (large) temporal extension $\mathcal{T}$ of the loop. 
In the (deformed) AdS space the expectation value of the loop is given by 
$\langle W(\mathcal{C}) \rangle \propto \exp [ i (S - S_0)]$, where $S$ is 
the Nambu-Goto action for the string configuration described above. 
$S_0$ is twice the Nambu-Goto action for a string hanging down from 
$z=0$ into the horizon at $z_h$. The latter contribution corresponds to 
the self energy of the quarks, and for moving quarks is closely related to 
the drag force acting on a single quark pulled through the plasma. 
The different AdS models discussed in the previous section 
enter the Nambu-Goto action. In the cases where the scalar $\Phi$ is a dilaton, 
the metrics entering there are not the ones given above, but need first to be 
multiplied by a factor involving the dilaton, for example 
for our choice of normalization 
$G^s_{\alpha\beta}= \exp(\sqrt{2/3}\Phi) G^E_{\alpha\beta}$. 
This gives rise to the respective metric in the string frame as necessary 
for the application of the Nambu-Goto action. 

One finds that there is always a maximal distance $L_{\rm max}$ 
between the quark and antiquark for which a string solution connecting 
them exists. At larger distances $L>L_{\rm max}$, no real-valued string solution 
can be found. For any smaller distance $L < L_{\rm max}$ there are two string 
solutions: The one coming closer to the horizon has a higher energy $F(L)$, 
while the one staying further away from the horizon has a lower energy $F(L)$. 
The former is thus unstable. In the following we want to 
concentrate on the maximal possible distance $L_{\rm max}$ 
that a bound quark-antiquark pair can have in the plasma. We call this 
distance the screening distance.\footnote{It must not be confused with the 
Debye screening length which governs the screening of the quark and 
antiquark when they are separated far beyond $L_{\rm max}$.} 
Interestingly, $L_{\rm max}$ is one of the observables that exhibit a 
universal behavior for large classes of non-conformal theories. 

\begin{figure}[ht]
  \centering
  \includegraphics[width=.56\linewidth]{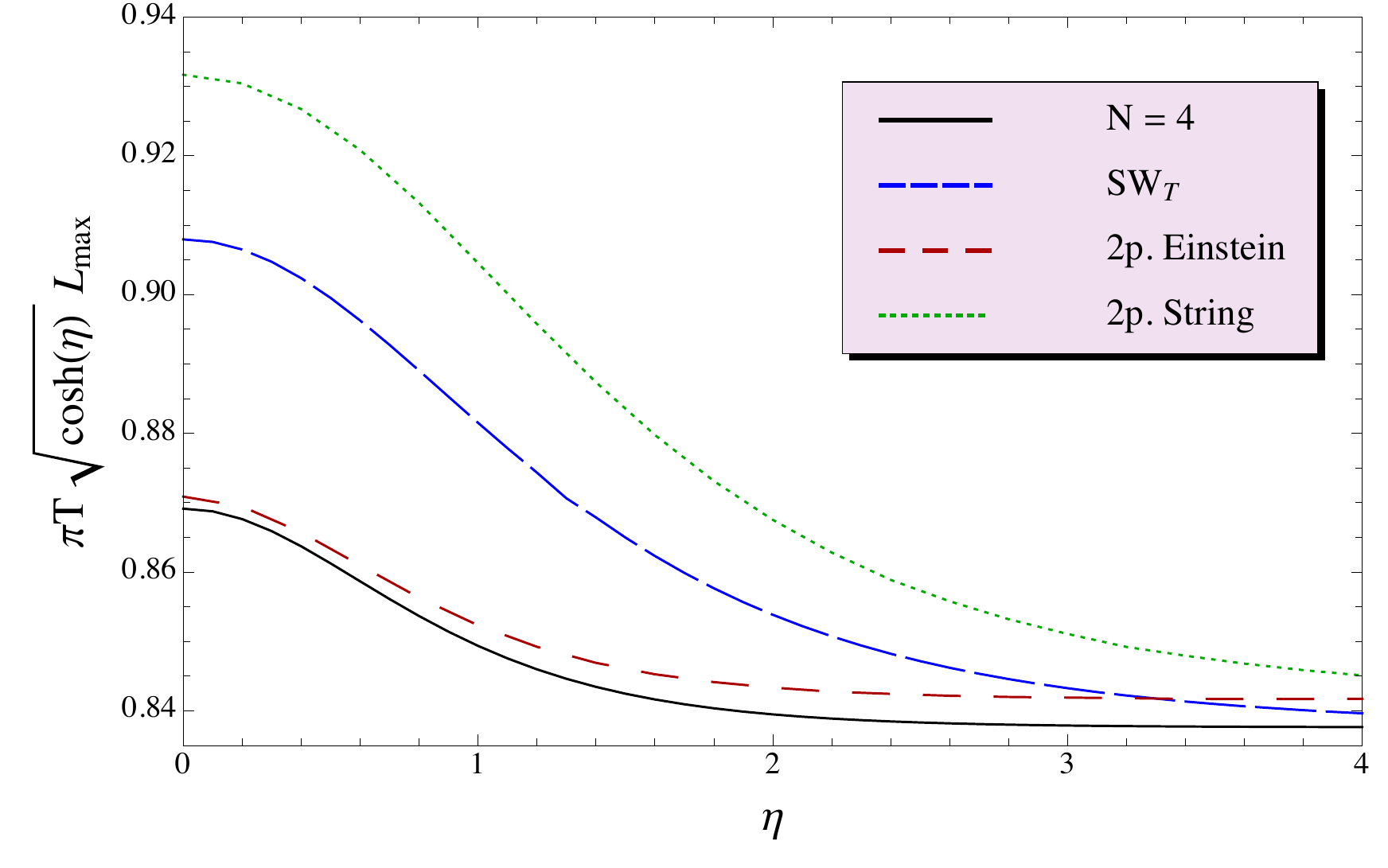}
  \caption{Dimensionless screening distance $\pi T L_{\rm max} \sqrt{\cosh \eta}$ 
of a $Q\bar{Q}$ pair as a function of rapidity $\eta$ for different models 
with vanishing chemical potential. 
  \label{fig1}}
\end{figure}

We have performed the calculation of $L_{\rm max}$ in all of the models 
presented in section \ref{sec:models}. Specifically, we have studied its dependence 
on the deformation parameters of the models and on all kinematical parameters, 
namely the velocity $v=\tanh \eta$ and 
the orientation angle of the pair with respect to its velocity in the plasma.
We first find that $L_{\rm max}$ is a robust quantity, i.\,e.\ it changes 
only relatively little with changes of the respective conformality-breaking 
parameters of the different models. For the case of vanishing chemical potential 
we have plotted in figure \ref{fig1} the dependence of $L_{\rm max}$ 
on the rapidity $\eta$ for several models with typical choices 
for their respective deformation parameters. 
The plot shows, for convenience, the dimensionless quantity $\pi T L_{\rm max}$, 
and in addition we have divided by the dominant $1/\sqrt{\cosh \eta}$ behavior 
of the $\eta$-dependence. 

For the case of vanishing chemical potential we have made the following 
interesting observation. For all given kinematical parameters 
the screening distance $L_{\rm max}$ is larger in any of the deformed 
models than in $\mathcal{N}\!=\!4$ SYM theory at the same temperature. 
It is very suggestive to conjecture that this minimality property of $L_{\rm max}$ 
in $\mathcal{N}\!=\!4$ SYM is more general and might even hold in 
all theories. 

\begin{figure}[t]
  \centering
  \includegraphics[width=.55\linewidth]{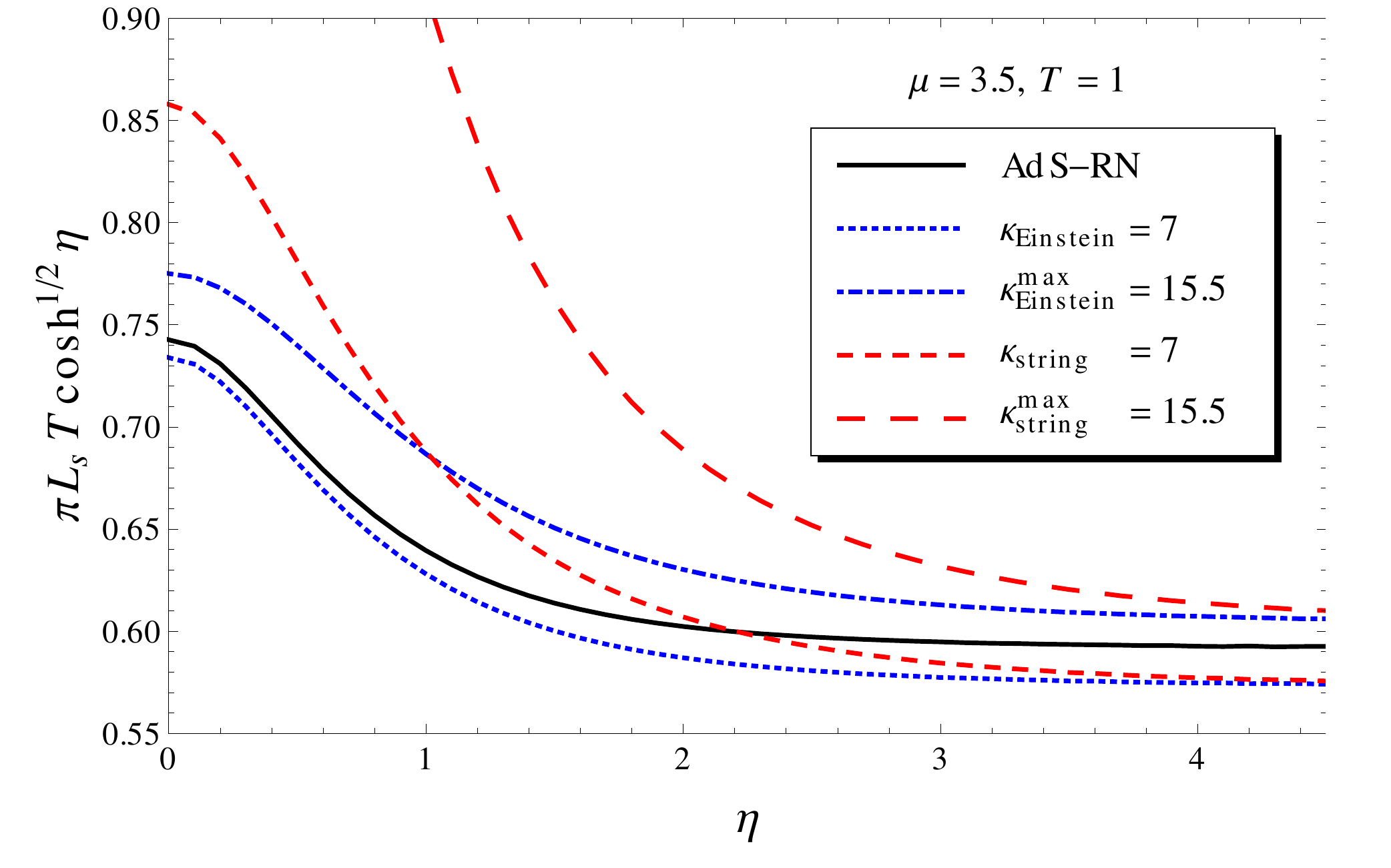}
  \caption{Dimensionless screening distance $\pi T L_{\rm max} \sqrt{\cosh \eta}$ 
of a $Q\bar{Q}$ pair as a function of rapidity $\eta$ for different models 
with finite chemical potential. 
  \label{fig:screeningDist1p}
}
\end{figure}

The $\eta$-dependence of $L_{\rm max}$ for finite chemical potential 
is shown in figure \ref{fig:screeningDist1p} for several models. 
Generally, the effect of changes in $\mu$ on $L_{\rm max}$ is smaller than that of 
changes in $T$. It turns out that for some of the Einstein-frame models and some 
range of deformation parameter $\kappa$ the screening distance becomes smaller than 
the corresponding value in $\mathcal{N}\!=\!4$ SYM at the same $T$ and $\mu$, 
thus violating the bound just discussed for the case of vanishing $\mu$. 
However, the violation stays always small. It is conceivable that an improved 
bound can be found for finite $\mu$. 

\begin{figure}[ht]
  \centering
  \includegraphics[width=.55\linewidth]{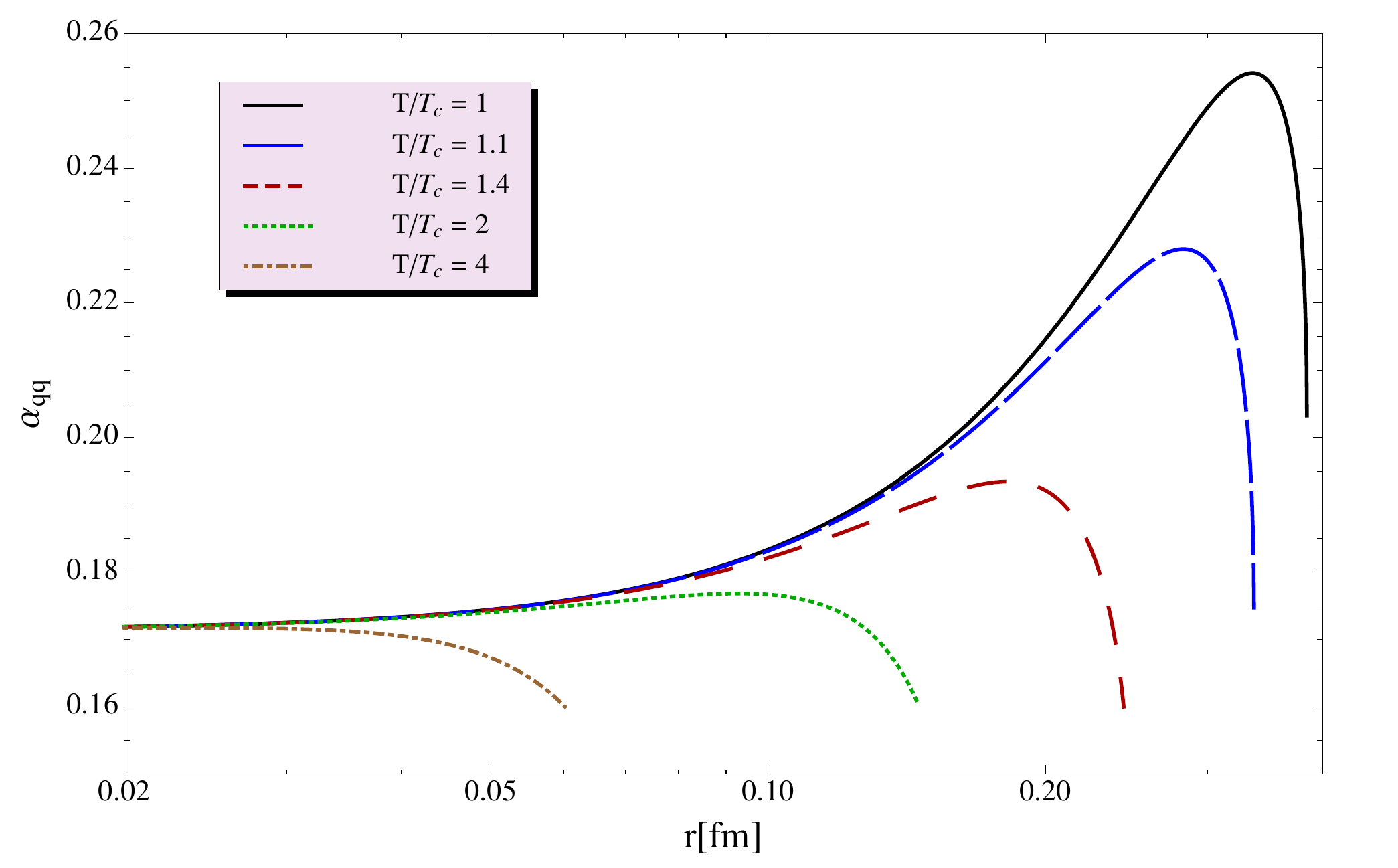}
  \caption{Running coupling $\alpha_{qq}$ as function of the distance $r$ for 
the temperatures $T/T_c$, $T/T_c = 1, 1.1, 1.4, 2$ and $4$ 
and for a fixed deformation $c$ in the SW${}_T$ model.
  \label{fig2}}
\end{figure}

Let us now turn to the running coupling in our AdS models. It can be defined 
in terms of the free energy $F(L)$ of the heavy quark-antiquark pair as 
\begin{equation}
\label{eq.ObsPhysQuant:RunCoup:CouplingDef}
\alpha_{qq} (r=L, \, T, \mu) \equiv \frac{3}{4} \, L^2 \, \frac{\tm{d} F(L, \, T, \mu)}{\tm{d} L} \,.
\end{equation} 
The normalization, here chosen as in QCD, 
is somewhat ambiguous as we do in general not 
know the Casimir factors of the theories dual to our AdS models. 
The normalization of $\alpha_{qq}$ further depends on 
the 't Hooft coupling of the theory. We therefore treat the normalization 
as a free parameter in our calculation. 
We have computed $\alpha_{qq}$ for various AdS models at vanishing 
and at finite chemical potential \cite{Schade:2012mqa,SambergDipl}. 
As an example of the $\mu=0$ case we show in figure \ref{fig2} 
the running coupling as obtained in the SW${}_T$ model for various 
temperatures. For $T_c$ we have chosen 176 MeV here. 
The curves end at the screening distance $L_{\rm max}$ of the respective 
temperature. We observe the universal property that $\alpha_{qq}$ 
develops a maximum before falling steeply close to $L_{\rm max}$. 
This holds for all non-conformal models, including those with 
finite chemical potential. Also for this observable, the effect of 
changing $T$ is considerably stronger than the effect of changing 
$\mu$. Let us finally point out that in the 2-parameter model 
at finite $T$ (but vanishing $\mu$) one can adjust the free parameters 
such that $\alpha_{qq}$ comes close to the corresponding QCD lattice 
data \cite{Kaczmarek:2004gv} over a sizeable range of temperatures. 

\section{Accelerated Motion}
\label{sec:accmotion}

Let us now turn to the accelerated motion of a heavy 
quark in the plasma. In the AdS/CFT framework, the simplest 
motion of this kind is a circular motion. 
The quark's energy loss due to radiation can, from a theoretical perspective, 
provide interesting information about the plasma, although 
a continuous circular motion of a quark is clearly not a realistic 
situation in any experiment. 

\begin{figure}[ht]
  \centering
  \includegraphics[width=.4\linewidth]{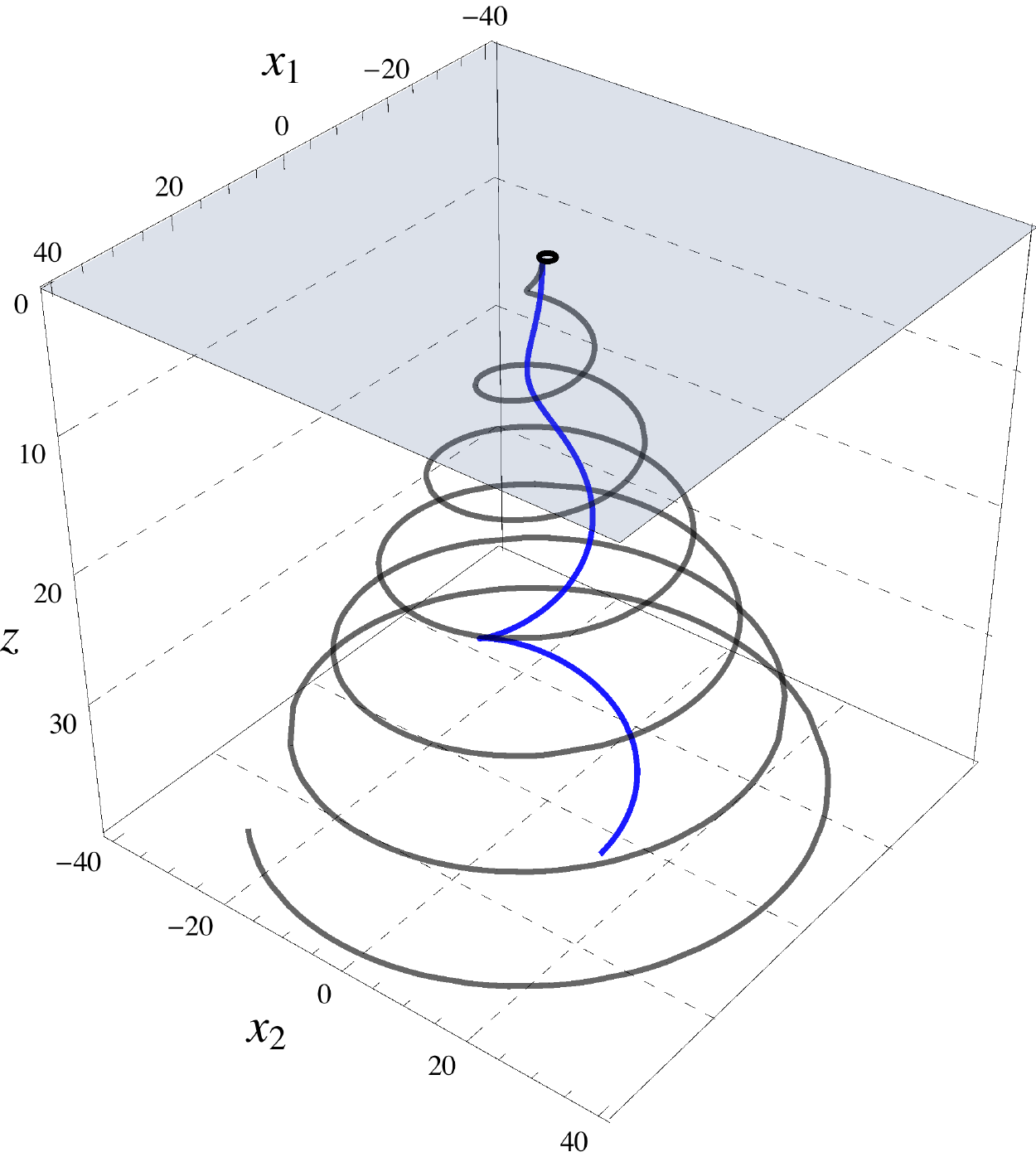}
  \hspace{2cm}
  \includegraphics[width=.4\linewidth]{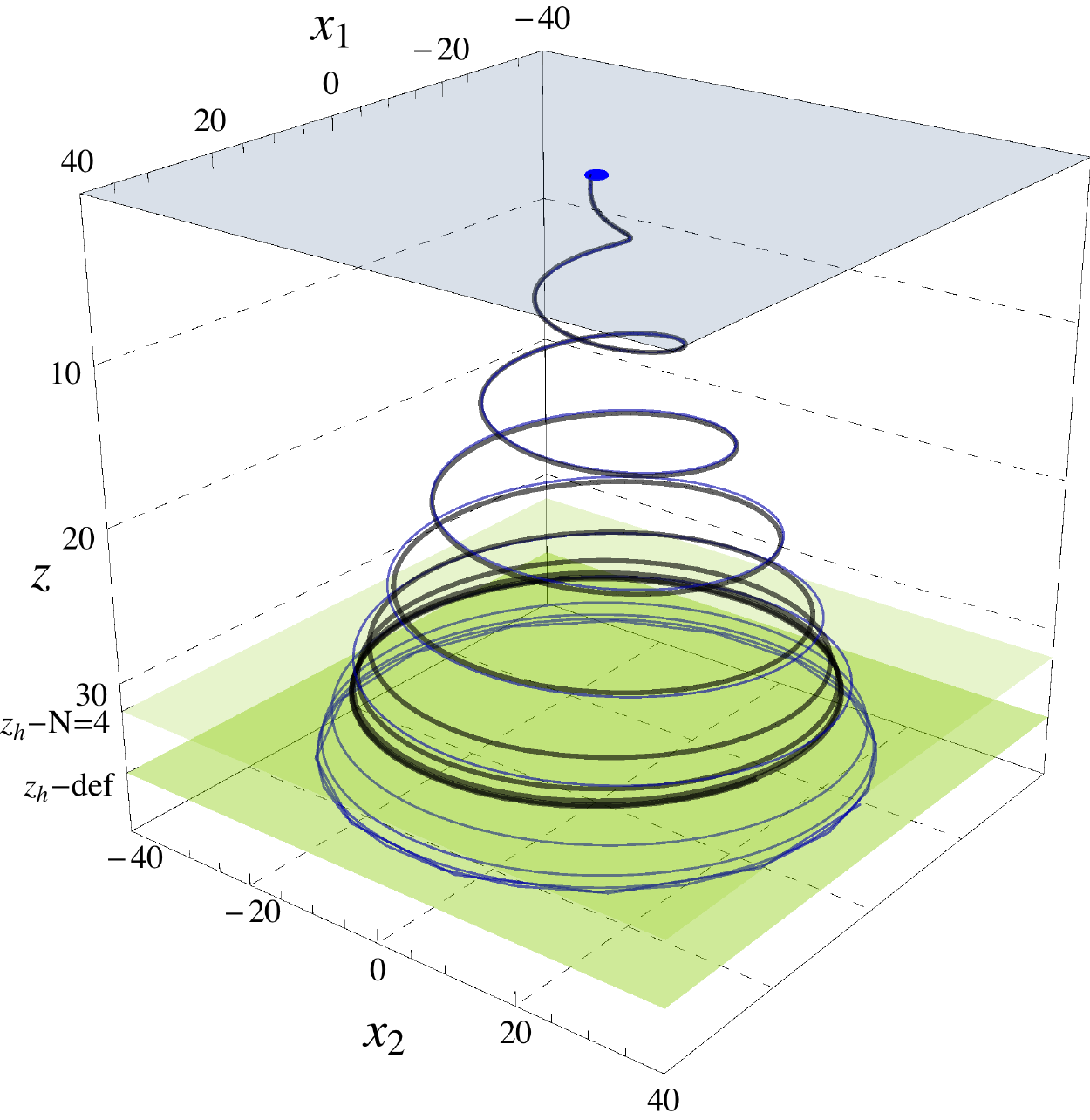}
  \caption{Left: String configuration in pure AdS${}_5$ 
($\mathcal{N}\!=\!4$ SYM at $T=0$) with 
radius $R_0 = 1$ and angular velocity $\omega = 0.3$ (blue) and $0.7$ (black). 
The quark rotates on the small black circle at the top. 
Right: String configuration in 2-parameter string-frame model (blue) for 
typical values of $\alpha$ and $c$ together with the corresponding 
$\mathcal{N}\!=\!4$ SYM case (black) at $T=0.01$, $R_0 = 1$ and $\omega = 0.7$ 
(in dimensionless units). 
  \label{fig3}}
\end{figure}

Let us consider a quark moving on a circle of radius $R_0$ with 
constant angular velocity $\omega$. For $\mathcal{N}\!=\!4$ SYM 
this situation has been considered for $T=0$ and $T\neq 0$ 
in \cite{Athanasiou:2010pv} and \cite{Fadafan:2008bq}, respectively. 
We have calculated the energy loss in the non-conformal metrics of 
section \ref{sec:models}, both at $\mu=0$ \cite{Schade:2012mqa}
and at $\mu \neq 0$ \cite{LinDipl}.
Typical configurations of the open string hanging down from the 
rotating quark into the bulk are shown in figure \ref{fig3}. 
The left panel corresponds to pure AdS${}_5$ space at $T=0$ and 
shows the string for two different angular velocities. For $T=0$ there is no black 
hole horizon and consequently the string spirals out to infinite radii 
deep in the bulk. 
The situation is different at finite $T$ where the motion of the string 
is limited in the holographic direction by the horizon, such that 
the string can reach only a finite radius, see right panel. 

We find that the string solution for the non-conformal models is 
always very close to the one for $\mathcal{N}\!=\!4$ SYM in the 
region close to the boundary, i.\,e.\ for small $z$. 
One can show that only the part of the string 
close to the boundary determines the energy radiation off the quark. 
Therefore, the radiation pattern in the non-conformal models is 
generally very similar to that of $\mathcal{N}\!=\!4$ SYM. 
This holds both at vanishing and finite chemical potential. 

\begin{figure}[ht]
  \centering
  \includegraphics[width=.5\linewidth]{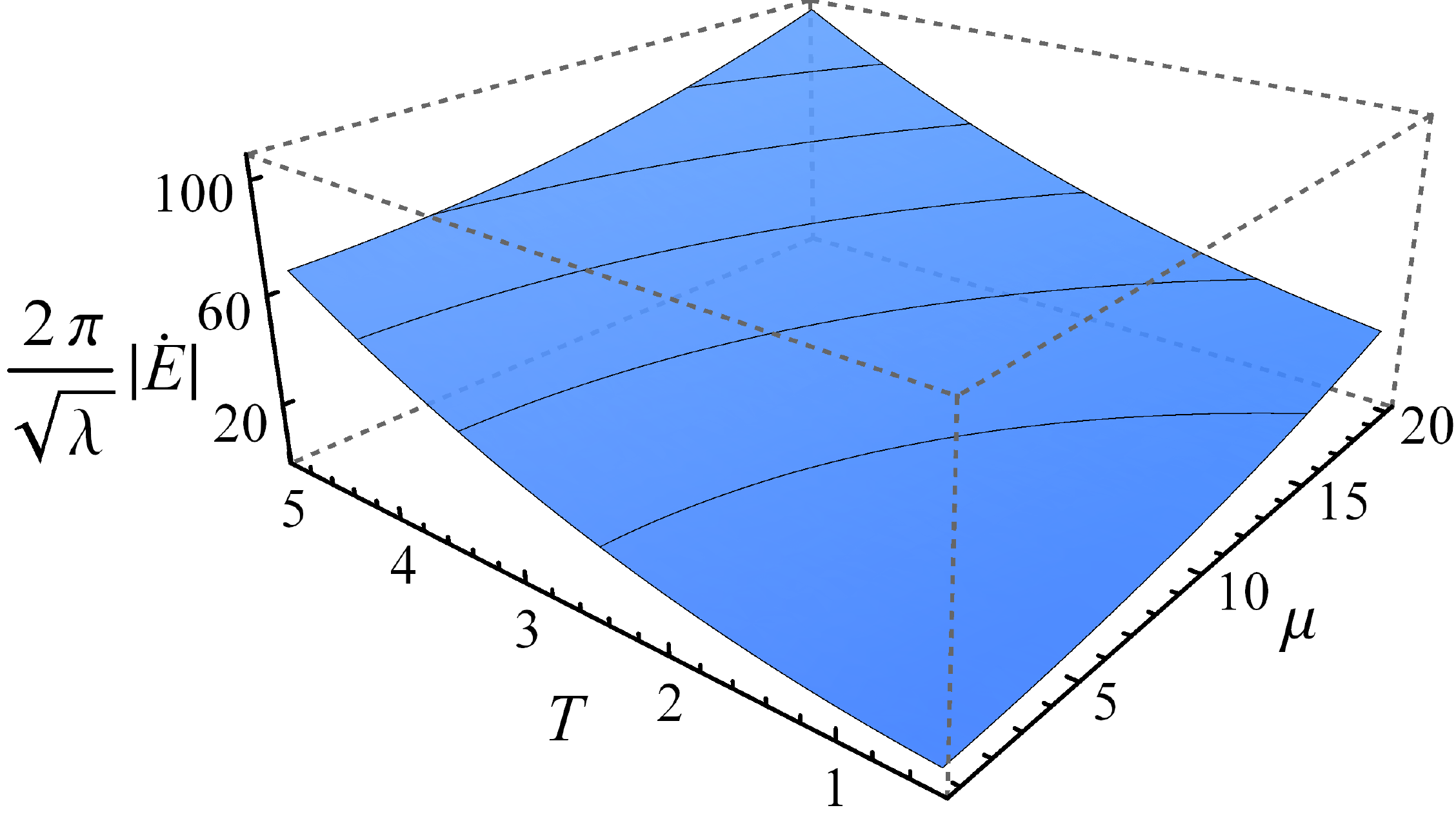}
  \caption{
Energy loss per unit time of a rotating quark in $\mathcal{N}\!=\!4$ SYM 
theory as a function of temperature $T$ and chemical potential $\mu$, 
both measured in the same arbitrary units. The quark moves with angular 
velocity $\omega = 1/2$ on a circle of radius $R_0 = 1$ in these units. 
  \label{fig4}}
\end{figure}

Finally, we observe that the rotating quark loses more energy at higher 
$T$ and $\mu$. Again, the effect of changing $T$ is considerably 
larger than the effect of changing $\mu$, as is seen in figure 
\ref{fig4} for the example of $\mathcal{N}\!=\!4$ SYM (without deformation) 
corresponding to an AdS-RN metric. 

\section{Summary}
\label{sec:summary}

We have studied several observables related to heavy quarks in 
strongly coupled plasmas in large classes of theories, using the framework 
of the AdS/CFT duality. The screening distance of a heavy quark-antiquark 
pair, the running coupling, as well as the energy loss of a heavy quark 
in circular motion have been studied. They are found to be 
robust under non-conformal deformations 
of the holographic theories and exhibit universal behavior. This 
universal behavior might be a sign of general strong-coupling 
dynamics, indicating a potential relevance of our findings for the 
quark-gluon plasma observed in heavy-ion collisions. 

\section*{Acknowledgments}

The work of K.\,S.\ was supported in part by the International Max Planck 
Research School for Precision Tests of Fundamental Symmetries. 
This work was supported by the Alliance Program of the
Helmholtz Association (HA216/EMMI).

\end{document}